\documentclass[amsmath,amssymb,aps,prl,superscriptaddress,twocolumn,longbibliography]{revtex4-2}

\usepackage{graphicx}
\usepackage{dcolumn}
\usepackage{bm}
\usepackage[usenames]{color}
\usepackage{comment}
\usepackage{hyperref}

\usepackage{multirow}
\usepackage[normalem]{ulem}

\begin{document}

\title{Attention-based Quantum Tomography }

\author{Peter~Cha}
\affiliation{Department of Physics, Cornell University, Ithaca, NY 14853, USA}
\author{Paul~Ginsparg}
\author{Felix~Wu}
\affiliation{Department of Computer Science, Cornell University, Ithaca, NY 14853, USA}
\author{Juan~Carrasquilla}
\affiliation{Vector Institute, MaRS Centre, Toronto, Ontario, M5G 1M1, Canada}
\affiliation{Department of Physics and Astronomy, University of Waterloo, Ontario, N2L 3G1, Canada}
\author{Peter~L.~McMahon}
\affiliation{School of Applied and Engineering Physics, Cornell University, Ithaca, NY 14853, USA}
\author{Eun-Ah~Kim}

\affiliation{Department of Physics, Cornell University, Ithaca, NY 14853, USA}
\begin{abstract}
With rapid progress across platforms for quantum systems,
the problem of many-body quantum state reconstruction for noisy quantum states becomes an important challenge.
There has been a growing interest in approaching the problem of quantum state reconstruction
using generative neural network models.
Here we propose the ``Attention-based Quantum Tomography'' (AQT), a quantum state reconstruction using an attention mechanism-based generative network that learns the mixed state density matrix of a noisy quantum state.
AQT is based on the model proposed in ``Attention is all you need" by Vaswani, et al. (2017) that is designed to learn long-range correlations in natural language sentences and thereby outperform previous natural language processing models.
We demonstrate not only that AQT outperforms earlier neural-network-based quantum state reconstruction on identical tasks but that AQT can accurately reconstruct the density matrix associated with a noisy quantum state experimentally realized in an IBMQ quantum computer.
We speculate the success of the AQT stems from its ability to model quantum entanglement across the entire quantum system much as the attention model for natural language processing captures the correlations among words in a sentence.
\end{abstract}
\maketitle

With rapid progress in modern quantum devices \cite{arute2019quantum}, the characterization and validation of large quantum systems becomes an important challenge.
Quantum state tomography offers a comprehensive characterization of quantum systems \cite{paris2004quantum}.
However, the exponential-in-$N_q$ Hilbert space of $N_q$-qubit many-body states implies that exact tomography techniques, such as Gaussian maximum likelihood estimation (MLE) \cite{james2001mle}, require exponential-in-$N_q$ amount of data as well as an exponential-in-$N_q$ time for processing.
Such prohibitive costs limit exact density matrix reconstruction to small system sizes $N_q\lesssim10$.
In fact, the tomographic measurement method that is integrated into IBM's Qiskit library is limited to $N_q=3$.
Hence,  many experiments rely on indirect methods of error determination, for example variants of randomized benchmarking~\cite{sheldon2016characterizing}. 
Indeed there are efforts to directly estimate properties of quantum states from measurements showing promising scaling~\cite{huang2020shadow}.
Nevertheless, the scalability of this approach depends crucially on the availability of global entangling gates acting on all qubits simultaneously, which are outside the reach of experimental systems.
Thus, new strategies for the characterization of noisy, entangled many-body quantum states using experimentally realistic measurements are much needed.

Recently, there has been a rapidly growing interest in using machine learning tools, such as deep neural networks, for quantum state reconstruction through generative modeling \cite{torlai2018neural, carrasquilla2019reconstructing, carrasquilla2019transformer}.
The foundation for this approach was laid in Ref~\cite{carleo2017solving}, which trained a restricted Boltzmann machine to represent complex quantum many-body states without requiring exponentially many parameters or memory size. However, the expressibility of restricted Boltzmann machines and scalability of training is typically restricted to pure, positive quantum states~\cite{carleo2017solving, bernien2017rydberg, torlai2018neural, torlai2019, devlugt2020}, single quantum oscillators \cite{tiunov2020}, and small mixed states~\cite{Torlai2018Latent}, which limits their applicability at the scale of modern noisy quantum computers. In contrast, Ref.~\cite{carrasquilla2019reconstructing} demonstrated, using a recurrent neural network (RNN), that generative neural network models trained on informationally complete positive operator-valued measurements (IC-POVM) may be capable of providing a classical description of a noisy quantum many-body state. (See Supplement ``Informationally complete positive operator-valued measurements''  for a brief introduction to the POVM formalism and a discussion of the POVM employed in this work) However, RNN-based tomography has so far only been demonstrated on classically simulated data, and despite promising indications, its ability to reconstruct a full density matrix has not been demonstrated even in simulation.

\begin{figure*}
\includegraphics[width=\linewidth]{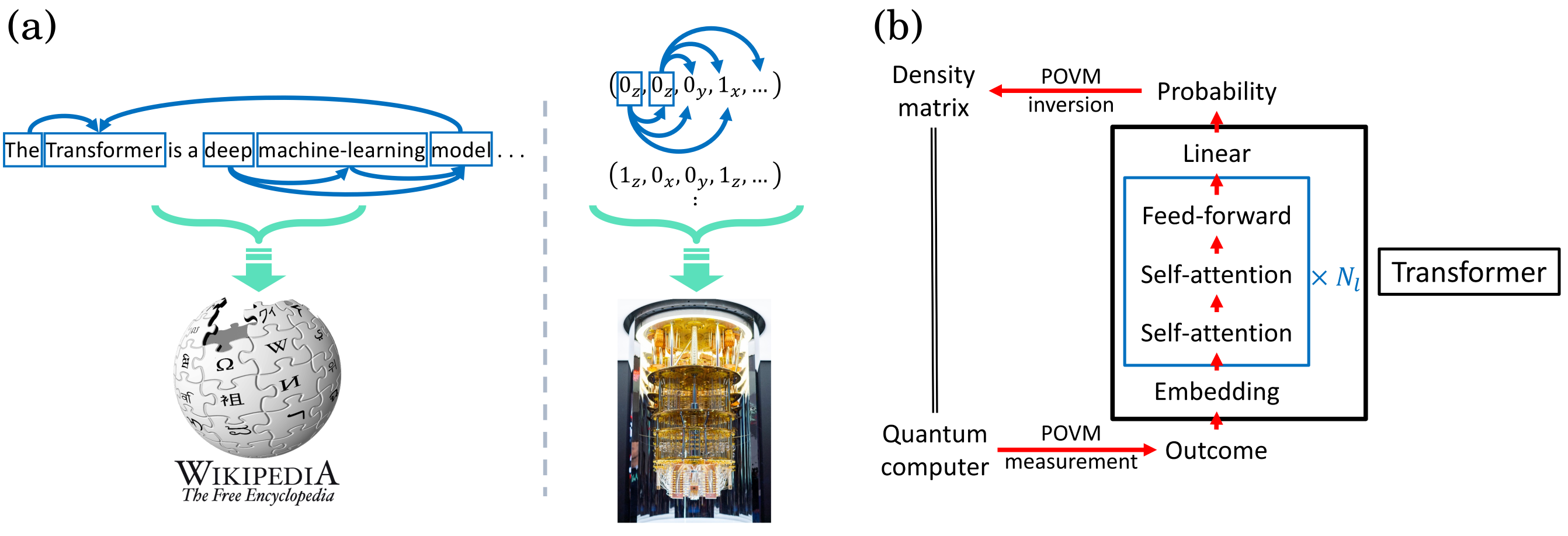}
\caption{(a) We illustrate the self-attention mechanism, which is key to the success of the Transformer architecture.
Left: The Transformer describes the ``state'' of Wikipedia after self-attention learns the correlations among words in the sentence specific to Wikipedia. Right: The AQT reconstructs the density matrix of a noisy mixed state (pictured: IBMQ) after self-attention learns the entanglement among qubits as it is reflected in projective measurements of the quantum state.
(b) A flowchart illustration of Attention-based Quantum Tomography (AQT), including a schematic diagram showing the internal structure of the Transformer neural network. In our work, we use $N_l=2$ layers and 64-dimensional embedding. (See Supplement ``Transformer neural network''  for a detailed discussion of the Transformer architecture in AQT)
}\label{fig1}
\end{figure*}

The Attention-based Quantum Tomography (AQT) adapts the Transformer architecture, a generative neural network model recently developed for natural language processing tasks~\cite{Vaswani2017}, for the task of quantum state tomography.
We begin by giving an intuition behind the Transformer and the rationale for its suitability for the tomography task.
We then demonstrate that AQT outperforms a previous RNN-based approach by a significant reduction in the sample complexity of the reconstruction procedure. We also simulate a simple faulty-qubit model and demonstrate the promise of AQT in the task of mixed-state reconstruction. Next, we deploy AQT on experimental data from IBMQ's quantum computer, showing strong qualitative agreement with MLE.
Finally, we demonstrate reconstruction of a density matrix with a system size that exceeds the reach of the tomographic tools offered publicly by IBMQ. \footnote{The implementation of AQT as well as all data used for this work are available at the public repository \url{https://github.com/KimGroup/AQT.}}

The rationale behind the AQT is our observation of a promising parallel between the task of natural language processing (NLP) and quantum state tomography (see Fig.~\ref{fig1}a). Sentences in natural language are highly structured with long-range relationships among their constituent words. Learning a language with an NLP model is the task of learning such structures and relationships by training on a set of sample sentences that constitute an extraordinarily tiny fraction of the complete set of all possible word combinations in the language. In more technical terms, this means training an autoregressive model to encode the conditional probabilities that govern which words may appear in which location in a sentence given the words that have come before it. In quantum state tomography, the key insight is that the density matrix representation of a quantum state is equivalent to the probability distribution of IC-POVM outcomes. Like sentences in natural language, entangled quantum states feature long-range correlations among their constituent qubits. The task of tomography is to learn this quantum state from a number of projective measurement outcomes that are as small as possible compared to the total space of projective measurement outcomes.

The Transformer~\cite{Vaswani2017}, which is the neural network architecture used in AQT, is an autoregressive model that employs the ``attention'' mechanism \cite{cho2014learning,cheng2016long,parikh2016decomposable}. It has been shown to be a dramatic step forward in efficiency and accuracy compared to previous state-of-art NLP models such as RNN~\cite{hochreiter1997long,chung2014empirical}. Before the Transformer, NLP tasks primarily relied on the RNN architecture~\cite{wu2016google,shazeer2017outrageously}, which incorporates correlations between words by passing an encoded ``memory'' of the words going back to the beginning of the sentence as each new word was read in sequentially. However, the correlations captured in this approach are inherently short-ranged, as the encoded memory in a sequential model such as the RNN suffers from exponential suppression in correlation~\cite{shen2019a}. The challenge of long-range correlations in semantic modeling were addressed with the Transformer architecture, which uses self-attention to study correlations between all words in a sentence simultaneously. As qubits in a many-body entangled quantum state have can have arbitrarily long-range entanglement, we can anticipate that the ability of the self-attention mechanism of the Transformer to capture long-range correlations among different positions of the data will be well-adapted to tomography.

As we schematically depict in Fig.~\ref{fig1}b, AQT takes as input a set of positive operator-valued measurement (POVM) outcomes, whether from a simulation or a real quantum device, and returns the reconstructed density matrix as output. The Transformer in AQT trains on a data set of $N_s$ one-shot local POVM outcomes $\{\vec{a}_i\}_{i=1}^{N_s}$ of a quantum state $\rho$, where each measurement is a vector $\vec{a}_i\in\{1,\ldots,N_a\}^{\otimes N_q}$ sampled from the distribution
\begin{align}
p_\rho(\vec{a})=\text{Tr}[M^{\vec{a}}\rho],\label{rho2p}
\end{align}
where $M^{\vec{a}}=(M^{a})^{\otimes N_q}$ are the operators defining the POVM. From this data, the Transformer learns a distribution $p_T(\vec{a})$ which is ideally close to $p_\rho$ and serves as a generative model that can sample from $p_T$ in linear-in-$N_q$ time. 
Once this training procedure is complete, \eqref{rho2p} can then be inverted for an appropriately chosen POVM. The POVM $T$-matrix is defined as $T^{\vec{a},\vec{a}'}=\text{Tr}\left[M^{\vec{a}}M^{\vec{a}'}\right]$. If $T^{\vec{a},\vec{a}'}$ is invertible, the reconstructed density matrix $\rho_T$ can be computed from the learned POVM distribution $p_T(\vec{a})$ in a post-processing step
\begin{align}
\rho=\sum_{\vec{a},\vec{a}'} p_T(\vec{a})T^{-1}_{\vec{a},\vec{a}'}M^{\vec{a}'}.\label{p2rho}
\end{align}
In this work, we use the Pauli POVM, which is invertible and easily accessible in the IBMQ quantum computers. \cite{carrasquilla2019reconstructing}

\begin{figure*}
\includegraphics[width=\linewidth]{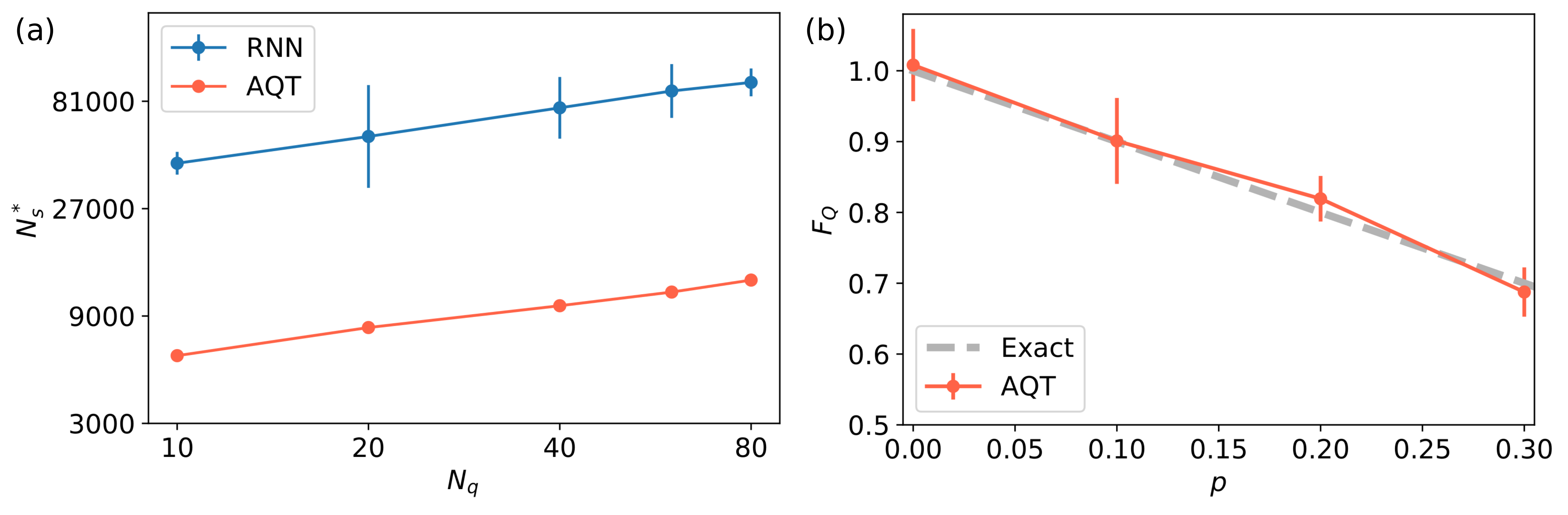}
\caption{(a) Log-log plot demonstrating scaling of necessary sample size $N_S^*$ for fixed classical fidelity $F_C=0.99$ in number of qubits $N_q$ using RNN \cite{carrasquilla2019reconstructing} and the AQT.
(b) Quantum fidelity $F_Q$ of states reconstructed using AQT, where each state was created with error rate $p$, which is the error parameter in the simulation to be characterized by the reconstruction. From $F_Q$ we can read off the error rate as $p=1-F_Q$. The expected result $F_Q=1-p$ is plotted in dashed grey.
}\label{fig2}
\end{figure*}

Our target state both in experiment and in classical simulations will be the $N_q$-qubit Greenberger-Horne-Zeilinger (GHZ) state
\begin{align}
|{\rm GHZ}\rangle_{N_q} = \tfrac{1}{\sqrt{2}}\bigl(|0\rangle^{\bigotimes N_q}+|1\rangle^{\bigotimes N_q}\bigr),
\label{eq:GHZ}
\end{align}
with system sizes ranging from $N_q=3$ to $90$ qubits. We choose the GHZ because it is a pure state of interest for quantum communication protocols and we can benchmark our results against others in the literature, including those that do not reconstruct the full density matrix \cite{carrasquilla2019reconstructing, huang2020shadow}.

A comprehensive measure of reconstruction is the quantum fidelity
\begin{align}
F_Q(\rho_{0},\rho_{1})=\left(\text{Tr}\left[\sqrt{\sqrt{\rho_{0}}\rho_{1}\sqrt{\rho_{0}}}\right]\right)^2,
\label{eq:FQ}
\end{align}
where $\rho_{0}$ is the target density matrix against which we compare the reconstructed density matrix $\rho_{1}$. Quantum fidelity in general requires full density matrix reconstruction \footnote{Recent work \cite{huang2020shadow} showed it is possible to estimate quantum fidelity without density matrix reconstruction in the special case when the target state is a pure state.}. We will carry out full density matrix reconstruction for small system sizes ($N_q\leq6$), for which we will be able to evaluate the exact quantum fidelity. However, in order to benchmark our results against the earlier works using neural networks, we initially investigate the classical fidelity, which can be used when the state reconstruction only yields measurement probabilities: 
\begin{align}
F_C(p_{0}, p_{1})=\sum_{\vec{a}}\sqrt{p_{0}(\vec{a})p_{1}(\vec{a})}.
\label{eq:FC}
\end{align}
Here the sum is over all IC-POVM outcomes $\vec{a}=(a_1,a_2,\ldots,a_{N_q})$, $a_i\in\{1,2,\ldots,N_a\}$ and $p_{0}$ and $p_{1}$ represent the measurement statistics of an IC-POVM over states $\rho_{0}$ and $\rho_{1}$, respectively. Even though the classical fidelity contains a number of terms exponential in $N_q$, it is possible to estimate $F_C(p_{0}, p_{1})$ efficiently by sampling from the generative model representing $p_{1}$, i.e.,  $F_C(p_{0}, p_{1})=\sum\limits_{\vec{a}}p_{1}(\vec{a})\sqrt{\frac{p_{0}(\vec{a})}{p_{1}(\vec{a})}}=\sum\limits_{\vec{a}\sim p_{1}}\sqrt{\frac{p_{0}(\vec{a})}{p_{1}(\vec{a})}}$, where the final sum is an average over $\vec{a}$ sampled from the distribution $p_1(\vec{a})$. This choice is enabled by the Transformer architecture, which allows for both exact sampling from $p_{1}$ and the exact calculation of $p_{1}(\vec{a})$ for any choice of $\vec{a}$ in time linear-in-$N_q$. However it should be noted that the classical fidelity only provides an upper bound on the quantum  fidelity~\cite{carrasquilla2019reconstructing}, and the discrepancy can be substantial \cite{huang2020shadow}. 

We first benchmark AQT against previous state-of-art neural quantum state tomography using RNN. Ref~\cite{carrasquilla2019reconstructing} studied an $N_q$-qubit GHZ state, with $N_q=10$ to $90$, using classically sampled measurements, and demonstrated that $N_s^*(N_q)$, the minimum size of training data for which the RNN can achieve a classical fidelity of 0.99, increases linearly with $N_q$. In Fig.~\ref{fig2}a, we demonstrate a similarly linear dependence of $N_s^*$ on $N_q$ using AQT, indicating that these natural language processing models can indeed learn non-trivial information about a quantum state from an amount of data that grows sub-exponentially with the system size. However, AQT exhibits an order-of-magnitude improvement in the sample complexity of learning the GHZ state compared to RNN tomography with a comparable slope $dN_s^*/dN_q$. As we will see shortly, this improvement in learning ability from RNN to AQT is even more dramatic in the task of density matrix reconstruction and quantum fidelity estimation, which is significantly more challenging than the task of achieving a good classical fidelity, and thus was out of reach for RNN tomography. \footnote{We note that the linear growth of sample complexity in system size for the classical fidelity threshold task seen in Fig. \ref{fig2}a does not translate to a linear growth in system size for a quantum fidelity threshold task.}

We now investigate the AQT's performance on a mixed state with a built-in simulated error. We consider a 3-qubit GHZ system and assume there is one faulty qubit, which we pick to be qubit-0. We assume that the faulty qubit flips ($0\leftrightarrow1$) with probability $p$. More precisely, this represents the mixed state
\begin{align}\label{eq:noisyqubit}
\rho_{err}(p)=(1-p)|{\rm GHZ}\rangle_3\langle {\rm GHZ}|_3+p|\psi\rangle_3\langle\psi|_3,
\end{align}
where $|\psi\rangle_3=\tfrac{1}{\sqrt{2}}\bigl(|100\rangle+|011\rangle\bigr)$.
For the small number of qubits that we study, we are able to compute the exact quantum fidelity. First, we consider the fidelity between the reconstructed state and the noisy state in Eq.~\ref{eq:noisyqubit}, for which we find  $F_Q(\rho_{\rm model}, \rho_{\rm err})=1$ within statistical error. This demonstrates that the AQT is sufficiently expressive to support a successful training procedure. To facilitate comparison to an experimental setting where $p$ is \textit{a priori} unknown, we compute the fidelity of the ``realized'' density matrix $\rho_{\rm model}$ to the ``target'' density matrix $\rho_{\rm GHZ}$, which is the error-free pure GHZ state. The numerical results for $p=0.0\sim0.3$ displayed in Fig.~\ref{fig2}b are consistent with the expectations from the built-in error. (See Supplement ``Error model \# 2'' for another model of error for the 3-qubit GHZ state) Note that as the density matrix reconstructed in AQT is not guaranteed to be positive, $F_Q\leq1$ is also not guaranteed. However, we find that the negative eigenvalues are less significant as the quality of reconstruction is increased. (See Supplement ``Positive density matrix reconstruction'' for a discussion of scaling of negative eigenvalues in sample size as well as one possible approach for positive density matrix reconstruction)

\begin{figure*}
\includegraphics[width=\linewidth]{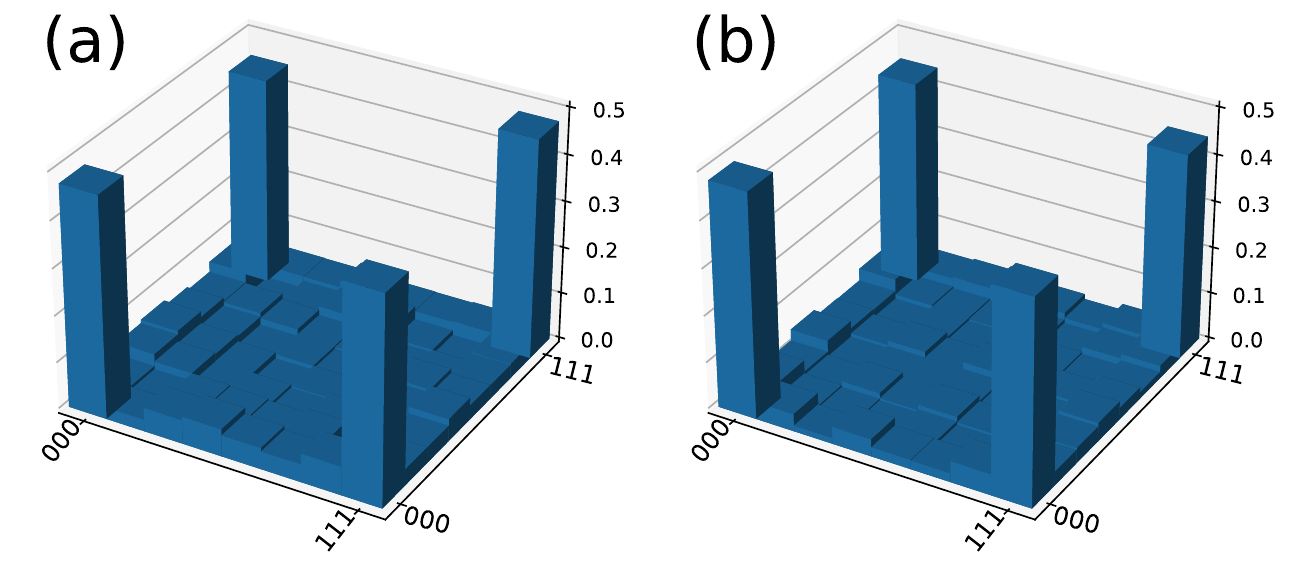}
\caption{Benchmarking AQT (a) to MLE tomography offered by IBM's Qiskit library (b) for a noisy 3-qubit QHZ state data generated on the IBMQ\_OURENSE quantum computer. Each bar represents the absolute value of a density matrix (DM) element.
}\label{fig:3qb}
\end{figure*}

Next, we benchmark the AQT aginst the MLE algorithm that is built into the IBM Qiskit library by performing tomography using the two approaches on the  measurements taken on IBMQ\_OURENSE on a 3-qubit system (Fig.~\ref{fig:3qb}).
For the reconstruction we took 100 measurements in each of $3^3=27$ possible measurement configurations, for a total of 2,700 measurements. Fig~\ref{fig:3qb} shows two reconstructed density matrices using the usual graphical representation.
Here, each bar represents a matrix element, in general complex, with the bar height set by its absolute value. The tall bar to the left is the density matrix element $|000\rangle\langle000|$, the bar to the rear is $|000\rangle\langle111|$, and so on. Note that the matrix elements represented by the bars in the rear and front are related by complex conjugation.
The AQT-reconstructed density matrix is in strong qualitative agreement with the MLE reconstruction, capturing the error in realizing the GHZ state on the quantum computer.
From the Transformer reconstruction, we find an exact quantum fidelity to the target pure GHZ state of $F_Q=0.917$, while the MLE reconstruction has fidelity $0.897$. \footnote{The quantum fidelity between the AQT and MLE reconstructions is 1.043.}
These results give a mutually consistent estimation of the reliability of the IBMQ\_OURENSE quantum computer.
The advantage of AQT as compared to exact tomographic methods such as MLE is that AQT can be scaled to larger systems.

\begin{figure*}
\includegraphics[width=0.6\linewidth]{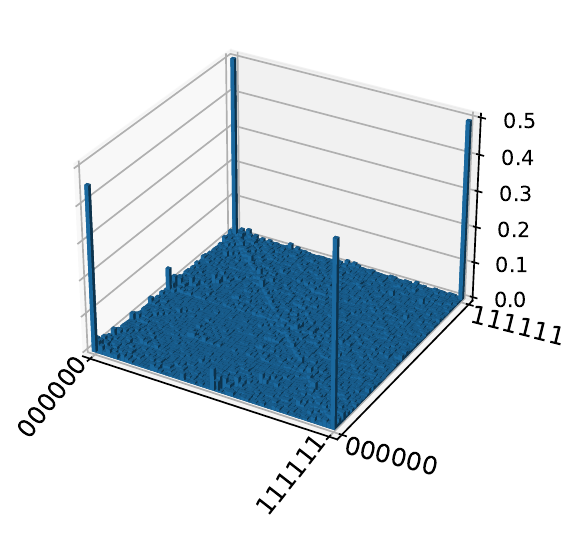}
\caption{\label{fig:6qb} Reconstructed density matrix of the 6-qubit GHZ state, using classically generated data. Each bar represents the absolute value of a matrix element.}
\end{figure*}

To further demonstrate the characterization ability of AQT, we reconstruct the density matrix for a 6-qubit GHZ state, already beyond the tomography functionality offered in Qiskit (see Fig.~\ref{fig:6qb}). We use classically generated data sampled from the noise-free GHZ state rather than data from IBMQ. (This is due in part to 
limited number of POVM measurements publicly accessible with IBMQ.) The reconstructed density matrix in Fig.~\ref{fig:6qb} uses a total of 200,000 measurements and has quantum fidelity 0.977. The reconstructed IC-POVM probability distribution $p_1$ (see Eq.~\eqref{eq:FC}) is in excellent agreement with the GHZ state, as expected from Fig.~\ref{fig2}a. Namely, achieving classical fidelity of $F_C=0.99$, which directly measures the accuracy of $p_1$ reconstruction, for a 6-qubit state requires even fewer than the 3,000 measurements required for the 10-qubit state.
On the other hand, the reconstruction of the full density matrix $\rho_1$ shows noise even with 200,000 measurements, though it is still in reasonable agreement with the GHZ state. (See Supplement ``Reconstruction of Dicke states'' demonstrating reconstruction of simulated 3- and 6-qubit Dicke states with 2700 and 72,900 measurements for comparison)
In general, an accurate reconstruction of $\rho_1$ requires much more data and computing time than an accurate reconstruction of $p_1$, since even small errors in $p_1$ are amplified into large errors in $\rho_1$. This is a restatement of the well-known fact that classical fidelity is an upper bound on quantum (exact) fidelity.
Exact error-scaling analysis in the number of samples is in general NP-hard \cite{suess2016} and remains an open question in AQT. The scaling of the POVM probability mean-squared error (MSE) in sample size $N_s$ for the 6-qubit GHZ state suggests that the AQT error scaling is comparable to statistical scaling MSE $\sim N_s^{-0.5}$, but with a significant reduction in overall magnitude compared to directly using the data. (See Supplement ``Transformer advantage'' for a detailed discussion of error scaling in sample size in AQT)

In summary, we  proposed the AQT which adopts elements of the Transformer, a generative deep neural network for NLP, to the task of quantum state tomography. The AQT outperformed earlier neural quantum tomography based on the RNN architecture
on an identical task, demonstrating a significant enhancement in the sample complexity of the reconstruction. This suggests that the AQT provides a nontrivial inductive bias suitable for the reconstruction of entangled states such as the ones considered in our experiments.  We constructed a qubit-error model and showed that AQT provides a reliable estimate of {\it a priori} unknown quantum mixed states and error rates in our specific setting. We then demonstrated for the first time that a machine learning based tomographic technique can reliably reconstruct the noisy density matrix of an entangled state, by reconstructing the 3-qubit GHZ state realized by a quantum computer provided publicly by IBMQ. Furthermore, using AQT we have reconstructed a 6-qubit GHZ state, which is a tomography task of a size beyond the reach of the tomography functionality in IBM Qiskit's software.

To the best of our knowledge, AQT represents the first machine-learning based approach to successfully reconstruct density matrices describing the states produced in an experimentally realized quantum computer.
AQT offers a substantial improvement in sample complexity over next-leading neural-network based tomographic methods, and is demonstrably capable of reconstructing noisy, full-rank states from experimental measurements. AQT is competitive with MLE in arbitrary density matrix reconstruction at small system sizes, but is also capable of POVM state characterization at large system sizes. Furthermore, AQT is designed to work with experimentally realizable local POVM measurements, without requiring globally-acting gates which are inaccessible in current experimental systems. Finally, AQT requires no assumptions about the entanglement structure or purity of the state being reconstructed and is expressive enough to characterize arbitrary states. With these features, AQT represents not only a leap forward in the cooperation of machine-learning and near-term quantum computing, but a powerful tomographic method uniquely suited to bridging the gap between simulation and experiment in the emerging era of noisy, intermediate-scale quantum computing.

AQT holds much promise for future progress. Because AQT learns a POVM representation of the quantum state that can be used to compute operator expectation values, a future investigation of AQT as a platform for shadow tomography will be fruitful. This work has been largely based on the GHZ state, facilitating a comparison with previous works without full density matrix reconstruction. Nevertheless the AQT is not inherently limited to a special pure state, and  an examination of how $N_s^*$ scales with $N_q$ in states with more complex entanglement will provide much insight into machine-learning based tomography. Tests on a bigger experimental system and other architectures will help us determine the full scalability of the AQT.
Furthermore, whether the AQT approach can build on the initial insight from our elementary error model towards more sophisticated error modeling and assessment to complement gate-set tomography~\cite{blume2013robust,blume2017demonstration} would be also an interesting direction. 

{{\bf Acknowledgements:} We thank Roger Melko, Giuseppe Carleo, Giaccomo Torlai, Mikhail Lukin, Markus Greiner for useful discussions. PC was supported by DOE Office of Basic Energy Sciences, Division of Materials Science and Engineering under Award DE-SC0018946. FW, and E-AK are supported by NSF HDR-DIRSE award number OAC-1934714 and in part by the Cornell Center for Materials Research with funding from the NSF MRSEC program (DMR-1719875). JC acknowledges support from Natural Sciences and Engineering Research Council of Canada (NSERC),  the Shared Hierarchical Academic Research Computing Network (SHARCNET), Compute Canada, Google Quantum Research Award, and the Canadian Institute for Advanced Research (CIFAR) AI chair program.}

{\bf Author Contributions:} EK, PC, PG planned the AQT. FW and PC implemented the AQT. EK, JC, PC, and PG were involved in benchmarking AQT against RNN and other architectures. EK, PC, PG, and PM were involved in experimenting on IBMQ. All authors contributed equally to writing the manuscript.

{\bf Competing Interests:} The authors declare no competing interests.

\bibliography{transformer}

\end{document}